\documentclass[10pt]{article}
\usepackage{charter} % change font
\usepackage{fullpage}
\usepackage{titlesec}

\usepackage{amsmath,graphicx}
\usepackage{amsfonts} % blackboard math symbols
\usepackage{amssymb} % varnothing
\usepackage{bm} % italic bold
\usepackage[font=small,skip=2pt]{caption}
\usepackage{subcaption}
\def\argmin{\mathop{\mathsf{arg\,min}}} % Argument of a minimization

\def\lim{\mathop{\mathsf{lim}}} % limit

\def\ebm{{\bm{e}}}

\def\xbm{{\bm{x}}}

\def\ybm{{\bm{y}}}

\def\rbm{{\bm{r}}}

\def\varphibm{{\bm{\varphi}}}

\def\thetabm{{\bm{\theta}}}
\def\phibm{{\bm{\phi}}}

\def\Hbm{{\bm{H}}}

\def\Fbm{{\bm{F}}}
\def\Sbm{{\bm{S}}}

\def\xbmhat{{\widehat{\bm{x}}}}

\def\C{\mathbb{C}}
\def\R{\mathbb{R}}

\def\phibm{{\bm{\phi}}}
\def\phibmhat{{\bm{\hat{\phi}}}}
\def\mbm{{\bm{m}}}

\begin{document}
 
\title{Deep Image Reconstruction using Unregistered Measurements\\without Groundtruth}
%An Unsupervised Learning Framework for Joint Image Reconstruction and Registration
%
% Single address.
% ---------------
{\author{{Weijie Gan$^1$, Yu Sun$^1$, Cihat Eldeniz$^2$, Jiaming Liu$^3$, Hongyu An$^2$ and Ulugbek S. Kamilov$^{1, 3}$} \\
\emph{{\small $^1$ Department of Computer Science and Engineering, Washington University in St. Louis, St.~Louis, MO, USA}} \\ 
\emph{{\small $^2$ Mallinckrodt Institute of Radiology, Washington University in St. Louis, St.~Louis, MO, USA}} \\
\emph{{\small $^3$ Department of Electrical and Systems Engineering, Washington University in St.~Louis, St. Louis, MO, USA}} \\ 
\small\emph{email}: \texttt{\{weijie.gan, sun.yu, cihat.eldeniz, jiaming.liu, hongyuan, kamilov\}@wustl.edu}
}
}

\date{}

\maketitle

\begin{abstract}
One of the key limitations in conventional deep learning based image reconstruction is the need for registered pairs of training images containing a set of high-quality groundtruth images. This paper addresses this limitation by proposing a novel \emph{unsupervised deep registration-augmented reconstruction method (U-Dream)} for training deep neural nets to reconstruct high-quality images by directly mapping pairs of unregistered and artifact-corrupted images. The ability of U-Dream to circumvent the need for accurately registered data makes it widely applicable to many biomedical image reconstruction tasks. We validate it in accelerated magnetic resonance imaging (MRI) by training an image reconstruction model directly on pairs of undersampled measurements from images that have undergone nonrigid deformations.
 \end{abstract}

% \begin{keywords}
% Image reconstruction, deep learning, magnetic resonance imaging, deformable image registration.
% \end{keywords}

\begin{figure*} 
    \centering
    \includegraphics[width=\textwidth]
    {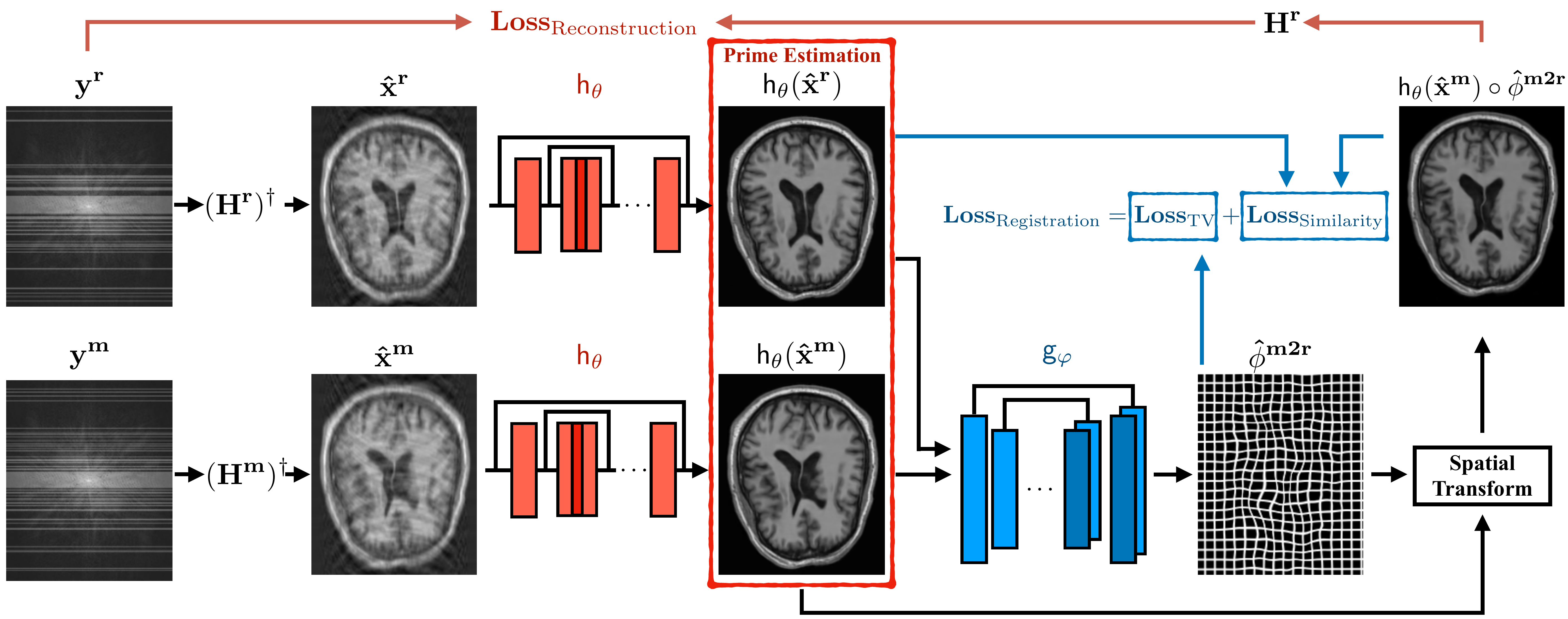}
    \caption{\textsc{U-Dream} jointly trains two CNN modules: $\mathsf{h}_\thetabm$ for image reconstruction and $\mathsf{g}_\varphibm$ for image registration, respectively. 
    Inputs are unregistered measurement pairs from the same subject. Their zero-filled images are passed through $\mathsf{h}_\thetabm$ to remove artifacts due to noise and undersampling. The output images are then used in $\mathsf{g}_\varphibm$ to obtain the motion field characterizing the directional mapping between their coordinates. We implement the wrapping operator as the Spatial Transform Network to register one of the reconstructed images to the other.}
    \label{fig:method}
\end{figure*}

\section{Introduction}
The reconstruction of a high-quality image $\xbm\in\C^n$ from a set of noisy measurements $\ybm\in\C^m$ is fundamental in biomedical imaging. 
For instance, it is essential in \emph{compressive sensing magnetic resonance imaging (CS-MRI)}~\cite{Lustig.etal2007}, which aims to shorten the acquisition time by obtaining diagnostic-quality images from severely undersampled k-space  measurements. 

\emph{Deep learning (DL)} has gained significant popularity in addressing the problem of image reconstruction in biomedical imaging~\cite{McCann.etal2017, Knoll.etal2020}.
A widely used strategy in this context trains a convolutional neural networks (CNN) to learn a mapping from the corrupted image to the desired high-quality image. 
Despite its success, the application of supervised DL can be challenging when it is difficult to collect a large number of high-quality training images. This limitation has motivated the research on \emph{unsupervised} DL schemes that rely exclusively on the information available in the corrupted data~\cite{Lehtinen2018, Krull.etal2019,Yaman2020}.
One widely used such technique is \emph{Noise2Noise}~\cite{Lehtinen2018} that trains a CNN by mapping pairs of observations of the same image containing different noise realizations.

Obtaining multiple accurately calibrated measurements from the same subject is a significant practical limitation. Subject motion during acquisition leads to unexpected structural deformations in the image. Failure to accommodate for such deformations complicates the training of DL methods. In this paper, we address this issue by proposing a novel \emph{\textbf{u}nsupervised \textbf{d}eep \textbf{re}gistration-\textbf{a}ugmented reconstruction \textbf{m}ethod (\textsc{U-Dream})}\footnote{We use the term \emph{unsupervised} to indicate the fact that groundtruth targets are not required during training.}. The novelty of our method is two-fold:
~\emph{(a)} \textsc{U-Dream} is an unsupervised DL scheme inspired from \emph{Noise2Noise} that learns directly from undersampled and noisy measurements $\ybm$. By mapping the reconstructed image back to the measurement domain, the method is trained to minimize the difference between the predicted measurements and the actual raw data, without using any fully-sampled groundtruth data.\ 
\emph{(b)} \textsc{U-Dream} simultaneously addresses the problems of registration and reconstruction by intergrating two separate CNN modules with different functionarities (See Fig.~\ref{fig:method}). The two CNNs are trained jointly by using pairs of unregistered measurements. 

\section{Background}
\label{sec-material}

Consider a problem of recovering an unknown image $\xbm$ from its linear measurements $\ybm = \Hbm\xbm + \ebm$, where $\ebm\in\C^m$ is the noise vector and $\Hbm\in\C^{m\times n}$ is the measurement operator. In traditional supervised DL, one learns a CNN model $\mathsf{h}_\thetabm$ that computes an inverse mapping from $\ybm$ to $\xbm$ by mitigating imaging artifacts and noise. It is common to perform this learning in the image domain by first computing an approximate inverse $\Hbm^\dagger$ of the measurement operator:~ $\xbmhat = \mathsf{h}_\thetabm(\Hbm^\dagger\ybm)$~\cite{McCann.etal2017, Knoll.etal2020}. For example, the measurement operator in CS-MRI can be represented as $\Hbm = \Sbm \Fbm$, where $\Fbm$ is the Fourier transform and $\Sbm$ is the k-space sampling operator. The mapping to the image space is performed by applying the zero-filled inverse Fourier transform: $\Hbm^\dagger = \Fbm^{-1}$.

\begin{figure*}
    \centering
    \includegraphics[width=\textwidth]{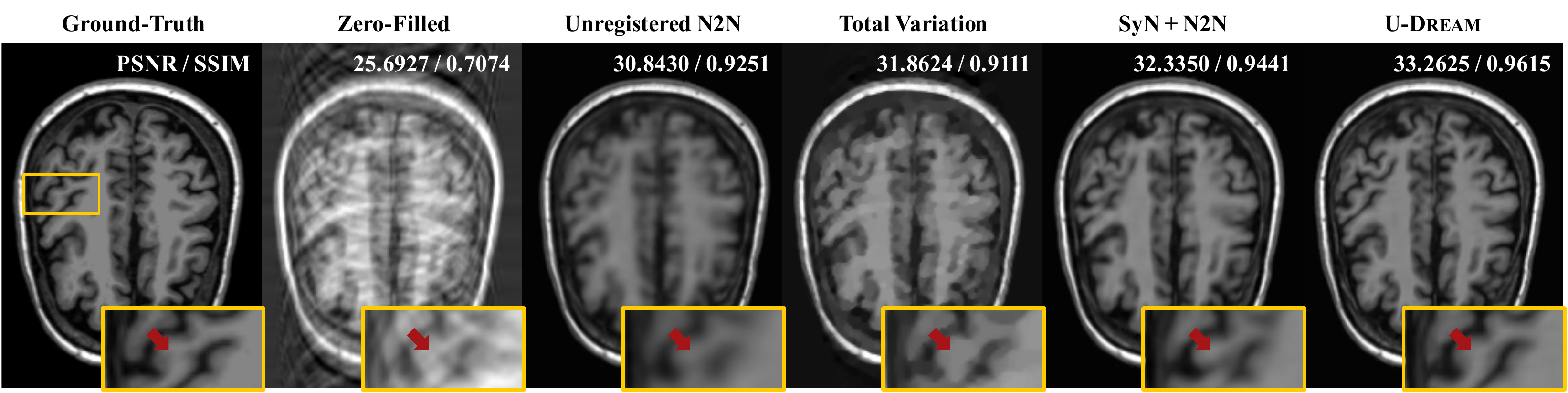}
    \caption{Illustration of several reconstruction methods on undersampled brain MRI data with an undersampling rate of 25\% and the deformation parameter of $\sigma=10$. The top-right corner of each image provides the PSNR and SSIM values with respect to the groundtruth image. \emph{Unregistered N2N} is directly trained on unregistered data, while \emph{SyN+N2N} trains the CNN on pre-registered but artifact-corrupted images. \textsc{U-Dream} achieves a significant improvement relative to both methods by jointly addressing the reconstruction and the registration.}
    \label{fig-rec} 
\end{figure*}

\emph{Noise2Noise (N2N)}~\cite{Lehtinen2018} is a recent technique for reducing the dependence of DL on high-quality ground-truth. It considers a group of \emph{registered} noisy images $\{\xbmhat_{ij}\}$ where $j$ indexes different realizations of the same underlying image $i$. For example, $ij$ might denote the $j^{\text{th}}$ MRI acquisition of the Subject $i$, with each acquisition consisting of an independent sampling pattern and noise realizations. The CNN in N2N can be trained via the following minimization
\begin{equation}
    \label{n2n}
    \argmin_\thetabm \sum_{i,j,j'} L\big(\mathsf{h}_\thetabm(\xbmhat_{ij}), \ \xbmhat_{ij'}\big),
\end{equation}
where $\mathsf{h}_\thetabm$ is the CNN parametrized by $\thetabm \in \R^p$ and $L$ is the loss function. 
While it is possible to acquire multiple independent views of the same subject, it is difficult to ensure the alignment of images across scans due to motion. Methods such as \emph{Noise2Void (N2V)}~\cite{Krull.etal2019} enable training on a single noisy image, but require structural incoherence of noise, which is not possible in some applications (see Fig.~\ref{fig-rec}).

DL has also gained significant attention in the context of image registration~\cite{Fu2020}. Let $\rbm$ represent the reference image and $\mbm$ its deformed counterpart. Image registration aims to obtain the registration field $\phibmhat^{\mbm 2 \rbm}$ that maps the coordinates of $\mbm$ to those of $\rbm$ by comparing the imaging content. Recently,~\cite{Balakrishnan2019} has introduced an unsupervised deep learning framework for deformable medical image registration. It formulates the registration as a CNN $\mathsf{g}_\varphibm$ with parameters $\varphibm \in \R^k$ mapping an input image pair $\{\mbm, \rbm\}$ to a deformation field $\phibmhat^{\mbm 2 \rbm}=\mathsf{g}_\varphibm(\mbm, \rbm)$ used for registration. The CNN is trained on a set of image pairs $\big\{\mbm_i, \rbm_i\big\}$ by minimizing the loss
\begin{equation}
    \label{voxelmorph}
    \argmin_\varphibm \sum_i L_\mathsf{d}(\mbm_i \circ \phibmhat^{\mbm 2 \rbm}_i, \rbm_i) + L_\mathsf{r}(\phibmhat^{\mbm 2 \rbm}_i),
\end{equation}
where $\circ$ denotes the wrapping operator that transforms the coordinates of $\mbm_i$ based on the registration field $\phibmhat^{\mbm 2 \rbm}_i$. The term $L_{\mathsf{d}}$ penalizes discrepancy between $\mbm_i$ after transformation and its reference $\rbm_i$, while $L_{\mathsf{r}}$ regularizes the local spatial variations in the estimated registration field.

%We can now introduce \textsc{U-Dream} that enables the training of the CNN directly on artifact-corrupted and unregistered images by integrating registration and reconstruction modultes. 

\section{Proposed Method}

Consider a pair of unregistered measurements $\ybm^\rbm$ and $\ybm^\mbm$ acquired from the same subject. We will refer to these images as the reference and the moving measurement, respectively
\begin{equation}
    \label{equ:pro-fwd}
    \ybm^\rbm = \Hbm^\rbm \xbm^\rbm  + \ebm^\rbm, \quad \ybm^\mbm = \Hbm^\mbm \big( \xbm^\rbm \circ \phibm^{\rbm 2 \mbm} \big) + \ebm^\mbm,
\end{equation}
Here, the subject motion is mathematically characterized as a non-rigid transformation field $\phibm^{\rbm 2 \mbm}$ in reference to $\xbm^\rbm$. One can then generate corrupted image pairs by simply applying the pseudoinverse of the respective forward operators
\begin{equation}
    \xbmhat^{\mbm} = {(\Hbm^{\mbm})}^\dagger\ \ybm^{\mbm}, \quad \xbmhat^\rbm = {(\Hbm^\rbm)}^\dagger\ \ybm^\rbm.
\end{equation}
As illustrated in Fig.~\ref{fig:method}, the proposed framework consists of two CNN modules, namely one  reconstruction module $\mathsf{h}_\thetabm$ and one registration module $\mathsf{g}_\varphibm$. 
$\mathsf{h}_\thetabm$ is a residual CNN~\cite{Lim2017} trained for reconstructing two clean images, $\mathsf{h}_\thetabm(\xbmhat^{\rbm})$ and $\mathsf{h}_\thetabm(\xbmhat^{\mbm})$, from $\xbmhat^{\rbm}$ and $\xbmhat^{\mbm}$, respectively. $\mathsf{g}_\varphibm$ is a CNN inspired from the widely-used UNet~\cite{Ronneberger2015} and trained to generate a motion field given the reconstructed image pair from $\mathsf{h}_\thetabm$: $\phibmhat^{\mbm 2\rbm} = \mathsf{g}_\varphibm\big(\mathsf{h}_\thetabm(\xbmhat^{\mbm}), \mathsf{h}_\thetabm(\xbmhat^{\rbm})\big)$.
Note that the input to $\mathsf{g}_\varphibm$ is order-sensitive and $\phibmhat^{\mbm 2\rbm}$ characterizes a directional mapping from the coordinate of $\mathsf{h}_\thetabm(\xbmhat^{\mbm})$ to the coordinate of $\mathsf{h}_\thetabm(\xbmhat^{\rbm})$.
\textsc{U-Dream} implements the differentiable wrapping operator $\circ$ as the \emph{Spatial Transform Network}~\cite{Jaderberg2015} to transform the coordinate of $\mathsf{h}_\thetabm(\xbmhat^{\mbm})$ and obtain a registered image $\mathsf{h}_\thetabm(\xbmhat^{\mbm}) \circ \phibmhat^{\mbm 2\rbm}$  with respect to $\mathsf{h}_\thetabm(\xbmhat^{\rbm})$.

The two CNNs are trained on a set of uncalibrated measurement pairs $\big\{\ybm^\rbm_i,\ybm^\mbm_i \big\}_i$. Additionally, we augment the training dataset by reversing the order of input pairs of $\mathsf{g}_\varphibm$ to obtain registration fields also in the opposite direction $\phibmhat^{\rbm 2 \mbm} = \mathsf{g}_\varphibm\big(\mathsf{h}_\thetabm(\xbmhat^{\rbm}), \mathsf{h}_\thetabm(\xbmhat^{\mbm})\big)$.

The loss function of the registration module consists of two components. The first component $L_\mathsf{d}$ penalizes difference of similarity based on local cross-correlation~\cite{Balakrishnan2019} and $L_\mathsf{r} = \lambda\ L_{\mathsf{TV}}$ imposes the total variation prior on the deformation field with parameter $\lambda > 0$
\begin{equation} \label{loss:regi}
    \begin{split}
        L_{\mathsf{reg}} = \sum_i\ L_{\mathsf{d}}\big(\mathsf{h}_\thetabm(\xbmhat_i^\rbm),\mathsf{h}_\thetabm(\xbmhat_i^\mbm) \circ \phibmhat^{\mbm 2\rbm}_i\big) + L_{\mathsf{r}}(\phibmhat^{\mbm 2\rbm}_i) +\ L_{\mathsf{d}}\big(\mathsf{h}_\thetabm(\xbmhat_i^\mbm),\mathsf{h}_\thetabm(\xbmhat_i^\rbm) \circ \phibmhat^{\rbm 2\mbm}_i\big) + L_{\mathsf{r}}(\phibmhat^{\rbm 2\mbm}_i).
    \end{split}
\end{equation}

On the other hand, the loss function of the reconstruction module is formulated in the measurement domain. It applies the forward operator on the reconstructed images after the alignment and penalizing the difference between the estimated and the measured k-space data:
\begin{equation}
    \label{loss:rec}
    \begin{split}
    L_{\mathsf{rec}} = \sum_i\ H\Big(\ybm_{i}^\rbm - \Hbm_{i}^\rbm\ \big(\mathsf{h}_\thetabm(\xbmhat_{i}^\mbm) \circ \phibmhat^{\mbm 2\rbm}_i\big) \Big) +\ H\Big(\ybm_{i}^\mbm - \Hbm_{i}^\mbm\ \big(\mathsf{h}_\thetabm(\xbmhat_{i}^\rbm) \circ \phibmhat^{\rbm 2\mbm}_i\big) \Big),
    \end{split}
\end{equation}
where $H$ is the Huber loss (or smoothed $\ell_1$ loss) defined as
\begin{equation}
    H(\xbm) = \begin{cases}
        0.5\xbm^2    &\text{if }|\xbm|<1 \\
        |\xbm| - 0.5 &\text{otherwise} \\ 
    \end{cases},
\end{equation}
which offers robustness to outliers.

Our training procedure alternatively minimizes the two loss functions by fixing the parameters of one CNN while training the other.

\section{Experiment}\label{sec-exp}

\noindent
\textbf{Setup.} 
The proposed method was quantitatively evaluated on accelerated MRI given 2D measurement pairs with non-rigid deformation. We used clean T1-weighted MR brain images from the open dataset OASIS-3~\cite{lamontagne2019oasis} by splitting the 60 subjects into 48, 6, and 6 for training, validation, and testing, respectively. For each subject, we extracted the middle 50 to 70 (depending on the shape of the brain) of the 256 slices on the transverse plane, containing the most relevant regions of the brain. We generated synthetic registeration fields by using the method in~\cite{Sokooti2017} and used them to deform the groundturth MR images. Three pre-defined parameters of the generation were the number of points randomly selected in the zero vector field $p=2000$, the range of random values assigned to those points $\delta=[-10, 10]$, and the standard deviations of smoothing Gaussian kernel for the vector field $\sigma\in\{10,18\}$. Thus, $\sigma$ is inversely related to the strength of deformation in the image. In order to obtain corrupted measurements, we simulated a single-coil MRI setting with a Cartesian under-sampling pattern (see Fig.~\ref{fig-rec}). We set the sampling rate to correspond to the 25\% of the complete k-space data and added the measurements noise corresponding to the input signal-to-noise ratio (SNR) of 40dB.

\medskip\noindent
\textbf{Comparisons.}
We compared \textsc{U-Dream} with four image reconstruction methods. 
The first method is the traditional total variation (TV) regularized image reconstruction~\cite{Beck2009, Liu2019}. The regularization parameter of TV was optimized for the best PSNR performance using the grid-search strategy.  The second method, called \emph{Unregistered N2N}, corresponds to the CNN trained by directly mapping the measurement pairs from unregistered images to one another. The third method, called \emph{Syn + N2N}, trains the CNN on the measurement pairs from pre-registered images. The images were registered by using \emph{Symmetric Normalization (SyN)}~\cite{avants2008symmetric}, which is one of the top-performing registration algorithms. Finally, \emph{N2V}~\cite{Krull.etal2019} was used to train the CNN on individual images without any registration. We used the peak signal-to-noise ratio (PSNR) and structural similarity index (SSIM) as image quality metrics.

\medskip\noindent
\textbf{Results.}
Fig.~\ref{fig-rec} shows visual results of image reconstruction for the deformation parameter $\sigma=10$. Among these results, zero-filled images contain ghosting and blurring artifacts. We have omitted the results of N2V due to its poor performance (comparable to ZF), which is expected as the artifacts in ZF are highly strucured. All other methods yield significant improvements over ZF. While \emph{Total Variation} shows considerable reduction in the aliasing artifacts, it leads to a loss of details due to the well-known ``staircasing effect.'' \emph{Unregistered N2N} method leads to a reasonable result even without registration in training, but it also contains a noticeable amount of blur, especially along the edges. While \emph{SyN + N2N} leads to a significant improvements over the traditional N2N and TV, it still suffers from smoothing in the region highlighted by the red arrow.
\textsc{U-Dream} outperforms all of these baselines methods in term of sharpness, contrast, and artifact-removal, which we attribute to its ability to jointly address registration and reconstruction. The quantitative results over the whole testing set is summarized in Table~\ref{tb-rec}, which highlights the significant quality improvements achieved by \textsc{U-Dream}.

\begin{table}
    \centering
    \caption{Quantitative evaluation of the reconstruction quality}
    \renewcommand\arraystretch{1.2}
    {\footnotesize
    \begin{tabular}{c|llll}
        \hline
        \textit{Schemes}       & \multicolumn{2}{c}{PSNR}                        & \multicolumn{2}{c}{SSIM}                        \\ \hline
        \textit{$\sigma$} & \multicolumn{1}{c}{10} & \multicolumn{1}{c}{18} & \multicolumn{1}{c}{10} & \multicolumn{1}{c}{18} \\ \hline
        Zero-Filled (ZF)       & 26.07                  & 26.02                  & 0.717                  & 0.715                  \\
        Unregistered N2N       & 29.03                  & 30.37                  & 0.903                  & 0.926                  \\
        Total Variation        & 29.78                  & 29.79                  & 0.893                  & 0.893                  \\
        SyN + N2N              & 30.31                  & 30.35                  & 0.929                  & 0.932                  \\
        \textsc{U-Dream}       & {\bf 31.60}            & {\bf 31.67}            & {\bf 0.945}            & {\bf 0.947}            \\ \hline
    \end{tabular}
    }
    \vspace{-1em}
    \label{tb-rec} 
\end{table}

\section{Conclusion}
\textsc{U-Dream} addresses an important problem of training a deep CNN directly from a set of unregistered artifact-contaminated images. It performs image reconstruction from unregistered measurements without any groundtruth by jointly performing  reconstruction and registration. We validated the method on undersampled MR measurement pairs corresponding to image pairs with non-rigid deformation. We observed that \textsc{U-Dream} leads to significant improvements compared to several baseline algorithms. Future work will further investigate \textsc{U-Dream} for other imaging modalities and datasets.

\section{Compliance with Ethical Standards}

This research study was conducted retrospectively using human subject data made available in open access by~\cite{lamontagne2019oasis}. Ethical approval was not required as confirmed by the license attached with the open access data.

\section{Acknowledgments}

Research reported in this publication was supported by the Washington University Institute of Clinical and Translational Sciences grant UL1TR002345 from the National Center for Advancing Translational Sciences (NCATS) of the National Institutes of Health (NIH).  The content is solely the responsibility of the authors and does not necessarily represent the official view of the NIH.

%%%%%%%%%%%%%%%%%%%%%%%%%%%%%%%%%%%%%%%%%%%%%
%% References
%%%%%%%%%%%%%%%%%%%%%%%%%%%%%%%%%%%%%%%%%%%%%


\begin{thebibliography}{10}

  \bibitem{Lustig.etal2007}
  M.~Lustig, D.~L. Donoho, and J.~M. Pauly,
  \newblock ``Sparse {MRI}: The application of compressed sensing for rapid {MR}
    imaging,''
  \newblock {\em Magn. Reson. Med.}, vol. 58, no. 6, pp. 1182--1195, December
    2007.
  
  \bibitem{McCann.etal2017}
  M.~T. McCann, K.~H. Jin, and M.~Unser,
  \newblock ``Convolutional neural networks for inverse problems in imaging: A
    review,''
  \newblock {\em IEEE Signal Process. Mag.}, vol. 34, no. 6, pp. 85--95, 2017.
  
  \bibitem{Knoll.etal2020}
  F.~Knoll, K.~Hammernik, C.~Zhang, S.~Moeller, T.~Pock, D.~K. Sodickson, and
    M.~Akcakaya,
  \newblock ``Deep-learning methods for parallel magnetic resonance imaging
    reconstruction: {A} survey of the current approaches, trends, and issues,''
  \newblock {\em IEEE Signal Process. Mag.}, vol. 37, no. 1, pp. 128--140, Jan.
    2020.
  
  \bibitem{Lehtinen2018}
  J.~Lehtinen, J.~Munkberg, J.~Hasselgren, S.~Laine, T.~Karras, M.~Aittala, and
    T.~Aila,
  \newblock ``{Noise2Noise: Learning image restoration without clean data},''
  \newblock in {\em ICML}, 2018, vol.~7, pp. 4620--4631.
  
  \bibitem{Krull.etal2019}
  A.~Krull, T.-O. Buchholz, and F.~Jug,
  \newblock ``{Noise2Void} - {L}earning denoising from single noisy images,''
  \newblock in {\em Proc. {IEEE} Conf. Computer Vision and Pattern Recognition
    ({CVPR})}, Long Beach, CA, USA, June 2019, pp. 2124--2132.
  
  \bibitem{Yaman2020}
  B.~Yaman, S.~A.~H. Hosseini, S.~Moeller, J.~Ellermann, K.~Ugurbil, and
    M.~Akcakaya,
  \newblock ``{Self-Supervised Physics-Based Deep Learning MRI Reconstruction
    Without Fully-Sampled Data},''
  \newblock in {\em Proc. Int. Symp. Biomedical Imaging}, 2020, pp. 921--925.
  
  \bibitem{Fu2020}
  Y.~Fu, Y.~Lei, T.~Wang, W.~J. Curran, T.~Liu, and X.~Yang,
  \newblock ``{Deep Learning in Medical Image Registration: A Review},''
  \newblock {\em Phys. Med. Biol.}, pp. 1--30, 2020.
  
  \bibitem{Balakrishnan2019}
  G.~Balakrishnan, A.~Zhao, M.R. Sabuncu, J.~Guttag, and A.~V. Dalca,
  \newblock ``{VoxelMorph: A Learning Framework for Deformable Medical Image
    Registration},''
  \newblock {\em IEEE Trans. Med. Imaging}, vol. 38, no. 8, pp. 1788--1800, 2019.
  
  \bibitem{Lim2017}
  B.~Lim, S.~Son, H.~Kim, S.~Nah, and K.~M. Lee,
  \newblock ``{Enhanced Deep Residual Networks for Single Image
    Super-Resolution},''
  \newblock in {\em CVPRW}, 2017, vol.~64, pp. 1468--1469.
  
  \bibitem{Ronneberger2015}
  O.~Ronneberger, P.~Fischer, and T.~Brox,
  \newblock ``{U-net: Convolutional networks for biomedical image
    segmentation},''
  \newblock in {\em MICCAI}, 2015, vol. 9351, pp. 234--241.
  
  \bibitem{Jaderberg2015}
  M.~Jaderberg, K.~Simonyan, A.~Zisserman, and K.~Kavukcuoglu,
  \newblock ``{Spatial transformer networks},''
  \newblock in {\em NIPS}, 2015, pp. 2017--2025.
  
  \bibitem{lamontagne2019oasis}
  LaMontagne \emph{et al.},
  \newblock ``{OASIS-3: longitudinal neuroimaging, clinical, and cognitive
    dataset for normal aging and Alzheimer disease},''
  \newblock 2019,
  \newblock medRxiv 2019.12.13.19014902.
  
  \bibitem{Sokooti2017}
  H.~Sokooti, B.~de~Vos, F.~Berendsen, B.~P.~F. Lelieveldt, I.~I{\v{s}}gum, and
    M.~Staring,
  \newblock ``{Nonrigid Image Registration Using Multi-scale 3D Convolutional
    Neural Networks},''
  \newblock in {\em MICCAI}, 2017, vol. 10433, pp. 232--239.
  
  \bibitem{Beck2009}
  A.~Beck and M.~Teboulle,
  \newblock ``{Fast gradient-based algorithms for constrained total variation
    image denoising and deblurring problems},''
  \newblock {\em IEEE Trans. Image Process.}, vol. 18, no. 11, pp. 2419--2434,
    2009.
  
  \bibitem{Liu2019}
  J.~Liu, Y.~Sun, C.~Eldeniz, W.~Gan, H.~An, and U.~S. Kamilov,
  \newblock ``{RARE: Image Reconstruction using Deep Priors Learned without
    Ground Truth},''
  \newblock {\em IEEE J. Sel. Top. Signal Process.}, 2020.
  
  \bibitem{avants2008symmetric}
  B.~B Avants, C.~L Epstein, M.~Grossman, and J.~C Gee,
  \newblock ``{Symmetric diffeomorphic image registration with cross-correlation:
    evaluating automated labeling of elderly and neurodegenerative brain},''
  \newblock {\em Med. Image Anal.}, vol. 12, no. 1, pp. 26--41, 2008.
  
  \end{thebibliography}
\end{document}